# How does stock market volatility react to oil price shocks?

Andrea Bastianin

University of Milan and FEEM

Matteo Manera

University of Milan-Bicocca and FEEM

Abstract: We study the impact of oil price shocks on the U.S. stock market volatility. We jointly analyze three different structural oil market shocks (i.e. aggregate demand, oil supply, and oil-specific demand shocks) and stock market volatility using a structural Vector Autoregressive model. Identification is achieved by assuming that the price of crude oil reacts to stock market volatility only with delay. This implies that innovations to the price of crude oil are not strictly exogenous, but predetermined with respect to the stock market. We show that volatility responds significantly to oil price shocks caused by unexpected changes in aggregate and oil-specific demand, while the impact of supply-side shocks is negligible.

Acknowledgments: We are grateful to two anonymous referees and Apostolos Serletis for insightful comments and suggestions. We also thank participants to: the International Workshop on "Oil and Commodity Price Dynamics" held at the Fondazione Eni Enrico Mattei (FEEM), Milan, 5-6 June 2014; the "8th International Conference on Computational and Financial Econometrics" held at the University of Pisa, 6-8 December 2014; the Conference on "Energy Markets" held at the IFP School-IFP Energies Nouvelles, Rueil-Malmaison, 17 December 2014; the "3rd International Symposium on Energy and Financial Issues (ISEFI 2015)" held at the IPAG Business School, Paris, 20 March 2015. The first author gratefully acknowledges financial support from the Italian Ministry of Education, Universities and Research (MIUR) research program titled "Climate change in the Mediterranean area: scenarios, economic impacts, mitigation policies and technological innovation" (PRIN 2010-2011, prot. n. 2010S2LHSE-001).

Corresponding author: Andrea Bastianin, Department of Economics, Management and Quantitative Methods (DEMM), University of Milan, Via Conservatorio 7, I-20122 Milan, Italy.

E-mail: andrea.bastianin@unimi.it

# 1. Introduction

In this paper we investigate the response of stock market volatility to oil price shocks. Policy makers, financial investors and journalists share the belief that unexpected changes in the price of crude oil can depress asset prices and boost volatility. Moreover, they tend to attribute the origin of oil price shocks mainly to oil production shortfalls due to political unrests in the Middle East and often consider the price of oil as exogenous with respect to macroeconomic and financial conditions.[1]

Conversely, most academics would agree that the price of crude oil is endogenous (Kilian, 2008b) and that it is driven by the combination of demand and supply side innovations (Hamilton, 2013).[2] However, the channels of transmission of energy price shocks and their impacts on macroeconomic and financial variables continue to be major topics for research and debate (see the survey by Kilian, 2014 and the discussion in Serletis and Elder, 2011).

---

[1] For the majority of financial investors and policy makers, the "prime suspects" for oil price run-ups are production disruptions due to political unrests in the Middle East (see e.g. Chisholm, 2014; Jakobsen, 2014; Kinahan, 2014; Saelensminde, 2014; Tverberg, 2010). Oil price shocks are associated with growth reductions (Jakobsen, 2014), inflationary pressures (Frisby, 2013; Saelensminde, 2014), debt defaults (Tverberg, 2010), systemic risk (Froggatt and Lahn, 2010), depressing effects on bond and stock markets (Frisby, 2013; Jakobsen, 2013; Regan, 2014; Saelensminde, 2014), as well as to volatility and uncertainty shocks (Froggatt and Lahn, 2010; Chisholm, 2014; Kinahan, 2014). For a more comprehensive view, which acknowledges the existence of shocks originating from both the supply and the demand side of the oil market, see The Economist (2012).

[2] See Blanchard and Galí (2009) and Blinder and Rudd (2013) for the alternative view that considers the price of oil as exogenous.

One example is the literature on the effects of oil price shocks on stock markets. Early analyses have two features in common: the price of oil is treated as exogenous and the causes underlying oil price shocks are not identified (see Chen et al., 1986; Huang et al., 1996; Jones and Kaul, 1996; Sadorosky, 1999; Wei, 2003). More recently, relying on the work of Kilian (2009), many studies have acknowledged that it is crucial whether a given oil price change has been generated by demand or supply pressures. In other words, the responses of stock prices (Abhyankar et al., 2013; Güntner, 2014; Jung and Park, 2011; Kilian and Park, 2009; Kang and Ratti, 2013a), dividend yield components (Chortareas and Noikokyris, 2014), and volatility (Degiannakis et al., 2014; Jung and Park, 2011) depend on the origin of the oil price shock. These results are not limited to the stock market. Actually, existing studies have confirmed that disentangling the causes underlying oil price shocks is important for explaining the response of many other variables, such as U.S. real GDP and inflation (Kilian, 2009), bond returns (Kang et al., 2014) and macroeconomic uncertainty (Kang and Ratti, 2013a,b). Moreover, these findings are not confined to the U.S., rather they hold also in international comparisons (see e.g. Abhyankar et al., 2013; Baumeister et al. 2010, Degiannakis et al., 2014; Güntner, 2014; Jung and Park, 2011; Kang and Ratti, 2013a; Kilian et al., 2009).

We build on the work of Kilian and Park (2009) to analyze the impact of oil price shocks on stock market volatility. Changes in the real price of crude oil are modeled as arising from three different sources: shocks to the supply of crude oil, to the aggregate demand for all industrial commodities and to oil-specific demand (also referred to as precautionary demand shocks). Kilian's (2009) structural Vector Autoregressive (VAR) model, used to describe the global market for crude oil, is augmented with a realized volatility proxy to investigate the response of the U.S. stock price volatility to oil price shocks. More precisely, we answer a number of questions. Does U.S. stock market volatility react to oil price shocks? Does the response depend on the source of the shock? What is the volatility response to oil price shocks for industry portfolios?

We show that U.S. stock market volatility has responded mostly to oil price shocks originating from the demand side. Positive shocks to aggregate demand cause an immediate reduction in the U.S. stock price volatility that lasts for about six months. Shocks due to unexpected increases in the precautionary demand for crude oil tend to boost U.S. stock market volatility, but only with delay. Supply side oil shocks have no impact on realized volatility, a result which is confirmed also using Kilian's (2008a) measure of exogenous oil supply shocks. Additional robustness checks show that our findings are not affected by changes to the sampling frequency of the data, or to the volatility proxy.

Consistently with the results obtained for U.S. stock price volatility, we show that the responses of the volatility of shares belonging to different industries vary depending on the cause underlying the oil price shock. On the contrary, only small differences do emerge in the timing and magnitude of the responses across industries.

Our study is related to the analyses of Jung and Park (2011) and Degiannakis et al. (2014). Jung and Park (2011) investigate the response of realized stock volatility in South Korea and Norway. For each country they estimate the global oil market model of Kilian (2009), augmented with the exchange rate and a measure of realized volatility. They find that the response of volatility to oil-specific demand shocks is positive and persistent if the country is an oil importer (i.e. South Korea), while it is not statistically significant for an oil-exporting country (i.e. Norway). Degiannakis et al. (2014) study the response of volatility to structural oil market shocks using the model by Kilian (2009). However, these authors focus on the European stock market, use a shorter sample period (1999-2010), and find that volatility reacts only to unexpected changes in aggregate demand, leaving no role for supply-side and oil-specific demand shocks.

The rest of the paper is organized as follows. Data and empirical methods are described in Section 2, while Sections 3 and 4 present the empirical results and some robustness checks. Section 5 concludes.

## 2. Data and empirical methods

### 2.1 Data

The volatility of the U.S. stock market is based on daily closing prices for the S&P500 index sourced from Yahoo! finance. However, since there are reasons to believe that industries might experience different reactions to oil price shocks, for instance because of heterogeneity in the level of energy intensity, we also consider a set of portfolios containing shares of firms in the same industry. For this part of the analysis, we rely on the data made available by Kenneth French, who provides daily returns for 49 industries.[3]

Realized volatility ($RV$) proxies the variability of the U.S. stock market. In line with Schwert (1989), $RV$ is calculated as the sum of the squares of daily real log-returns[4]:

$$RV_t = \sum_{k=1}^{N_t} r_{j:t}^2 \qquad (1)$$

where $N_t$ and $r_{j:t}$ are the number of days and daily real log returns in month $t$, respectively. All empirical results are based on the annualized realized standard deviation, defined as $(252 \times RV_t)^{1/2}$, although for brevity we keep on using $RV$ thereafter.

The relationship between oil market shocks and stock price volatility can be explored empirically based on a structural VAR model for $\mathbf{z}_t = [\Delta prod_t,\ rea_t,\ rpo_t,\ RV_t]^T$. Equations for the first three variables correspond to the structural VAR model of Kilian (2009). This model describes the global market for crude oil using the annualized percent change in world crude oil production, $\Delta prod_t$, an index of real economic activity, $rea_t$, and the real price of oil, $rpo_t$.[5] Data are monthly and the sample period runs from February 1973 until December 2013.

---

[3] See http://mba.tuck.dartmouth.edu/pages/faculty/ken.french/data library.html, for details.

[4] The construction of real returns on portfolios and on the S&P500 follows Lunde and Timmermann (2005). Since the Consumer Price Index, (CPI) is available at monthly frequency, we linearly interpolate monthly CPI data such that the resulting daily variable grows at constant rate through the month. The end-of-month observation of the daily CPI variable is thus equal to the corresponding value of the monthly CPI series. The price index used is the CPI for All Urban Consumers, as reported by the Bureau of Labor Statistics (mnemonic: CPIAUCSL).

**2.2 Identification and estimation of the impact of oil price shocks on volatility**

The structural VAR model used to estimate the response of U.S. stock price volatility to oil market shocks is recursively identified along the lines of Kilian (2009) and Kilian and Park (2009) and can be written as:

$$\mathbf{A}_0 \mathbf{z}_t = \boldsymbol{\alpha} + \sum_{i=1}^{24} \mathbf{A}_i \mathbf{z}_{t-i} + \boldsymbol{\varepsilon}_t \qquad (2)$$

Serially and mutually uncorrelated structural innovations, $\boldsymbol{\varepsilon}_t$, are derived from the vector of reduced-form VAR innovations, $\mathbf{e}_t = \mathbf{A}_0^{-1} \boldsymbol{\varepsilon}_t$, by imposing the following exclusion restrictions on $\mathbf{A}_0^{-1}$:

$$\mathbf{e}_t \equiv \begin{pmatrix} e_t^{\Delta prod} \\ e_t^{rea} \\ e_t^{rpo} \\ e_t^{RV} \end{pmatrix} = \begin{bmatrix} a_{11} & 0 & 0 & 0 \\ a_{21} & a_{22} & 0 & 0 \\ a_{31} & a_{32} & a_{33} & 0 \\ a_{41} & a_{42} & a_{43} & a_{44} \end{bmatrix} \begin{pmatrix} \varepsilon_t^{oil\ supply\ shock} \\ \varepsilon_t^{aggregate\ demand\ shock} \\ \varepsilon_t^{oil\text{-}specific\ demand\ shock} \\ \varepsilon_t^{other\ shocks\ to\ RV} \end{pmatrix} \qquad (3)$$

---

[5] $\Delta prod_t$, the annualized percent change in world crude oil production, is defined as $1200 \times \ln(prod_t/prod_{t-1})$. World oil production, $prod_t$, is available starting from January 1973 in the U.S. Energy Information Administration's Monthly Energy Review (Table 11.1b). The index of real economic activity, $rea_t$, introduced by Kilian (2009), is based on dry cargo ocean shipping rates and is available on the website of the author. It is used to proxy monthly changes in the world demand for industrial commodities, including crude oil. The real price of crude oil, $rpo_t$, is the refiner's acquisition cost of imported crude oil and it is available from the U.S. Energy Information Administration (EIA). Deflation is carried out using the CPI for All Urban Consumers, as reported by the Bureau of Labor Statistics (mnemonic: CPIAUCSL). The deflated price is in logarithms and then is expressed in deviations from its sample average.

We can think of Equation (3) as being composed of two blocks. The first three rows describe the global oil market block as in Kilian (2009), the second block (i.e. the last row) consists of one equation for the stock price volatility.

Exclusion restrictions in the first block are consistent with a global market of crude oil characterized by a vertical short-run supply curve and a downward sloping short-run demand curve. Oil supply does not respond within the month to changes in oil demand, but it shifts in response to changes in production due to exogenous events, such as conflicts in the Middle East. Oil demand is driven by the remaining structural innovations. Aggregate demand shocks capture shifts in the demand for all industrial commodities, including crude oil, associated with the global business cycle. The restrictions in the second row of Equation (3) imply that oil-specific demand shocks (also referred to as precautionary demand shock) influence the global business cycle only with a delay. The last structural shock in the first block (i.e. oil-specific demand innovations) is designed to capture changes in the real price of oil that are not explained by oil supply and aggregate demand shocks.[6] Therefore, the real price of oil changes instantaneously in response to both aggregate and oil-specific demand shocks, as well as in response oil supply shocks.

The structural innovations in the last row of Equation (3) are not true structural shocks, rather a residual category reflecting innovations to stock market volatility which are not driven by oil market shocks.

Our identification scheme relies on the additional assumption that innovations to oil production, real economic activity and the real price of crude oil are predetermined with respect to U.S. stock price volatility. In other words, while oil production, real economic activity and the real price of crude oil respond to all past information, predeterminedness implies the absence of an instantaneous feedback from *RV* to those variables. This working hypothesis has been used extensively in the literature (see Kilian 2008b and references therein, as well as Kilian and Park, 2009), and it is also empirically supported by the results of Kilian and Vega (2011).

---

[6] It is important to note that the oil-specific demand shocks are a residual category, rather than a structural "precautionary demand shock". For an example of an explicitly identified "speculative oil demand shock", see Kilian and Murphy (2014).

## 3. Empirical results

### 3.1 The impact of oil price shocks on the volatility of the U.S. stock market

One of the key results of Kilian (2009) is that, at each point in time, shocks to the real price of crude oil are the result of disturbances originating both from the supply and the demand sides of the market. For instance, the volatility of supply side innovations has decreased through time, and supply shocks seem to have no role in explaining the surge in the price of oil in 2008. This fact is at odds with the view shared by the majority of policy makers and financial investors, according to which a direct causal link between volatility and political events in the Middle East is often postulated, while little role, if any, is attributed to shocks arising from the demand side.[7]

**[FIGURE 1 ABOUT HERE]**

Responses of U.S. stock market volatility to a (one-standard deviation) shock to the supply and demand of crude oil are reported in Figure 1. Each panel shows the estimated impulse response function (IRF), together with one and two-standard error bands (namely, 68% and 95% confidence intervals), based on the recursive-design wild bootstrap of Gonçalves and Kilian (2004). Henceforth, oil price shocks will represent unpredictable reduction to the supply crude oil and unpredictable aggregate or oil-specific demand increases. In other words, all shocks have been normalized such that their expected effect is to generate an increase in the price of crude oil.

---

[7] A case in point is Kinahan (2014), who writes: "*the market's drop - triggered by higher oil prices and the potential for greater oil supply disturbances in Iraq - stirred investor risk perception. As evidence the CBOE Volatility Index,…, hit 12.56 on June 12*".

As it can been seen from a joint inspection of the plots in Figure 1, U.S. stock volatility has responded mostly to oil price shocks originating from the demand side of the oil market, while supply-driven shocks have had hardly any impact.

The first graph on the left shows that shocks to the supply of crude oil have no impact on the U.S. stock price volatility: the estimated impulse response function is never statistically distinguishable from zero. From the graph in the middle we see that an unanticipated increase of the aggregate demand for industrial commodities yields an immediate decrease in realized volatility, which is significant at the 68% confidence level. After about six months, the response of realized volatility becomes statistically indistinguishable from zero, while after a year the sign of the response switches from negative to positive but remains statistically significant for one month using the one-standard error bands.

The response of stock price volatility to a shock to the precautionary demand for crude oil is shown in the rightmost panel of Figure 1. Similarly to shocks to aggregate demand, the impact response of realized volatility to unexpected increases in oil-specific demand is negative. However, after a semester the response becomes positive and statistically significant at the 68% confidence level. A year after the shock the response is not statistically distinguishable from zero. The delayed and temporary volatility boosting effect of an unexpected increase in oil-specific demand could be explained by recalling that shocks to precautionary demand for oil are can be interpreted as shocks to the expectations about future oil supply. Therefore, a sustained higher precautionary demand could indicate higher macroeconomic uncertainty, which is clearly reflected in a more volatile stock market.

Overall, the three impulse response functions are consistent with the view that the origins of oil price shocks matter for explaining the response macroeconomic and financial variables (Abhyankar et al. 2013; Chortareas and Noikokyris, 2014; Degiannakis et al. 2014; Güntner, 2014; Kilian, 2009; Kilian and Park, 2009; Kang and Ratti 2013a,b; Kang et al., 2014). In the case of stock price volatility, this implies that, if investors know what has caused an increase in the price of crude oil, they can optimize their risk management and asset allocation strategies accordingly.

The fact that oil price increases caused by aggregate demand shock lead to a reduction of U.S. stock market volatility might seem counterintuitive at first. However, as pointed out by Kilian (2009) and Kilian and Park (2009), unexpected increases in aggregate demand have two simultaneous effects. On the one hand, they signal improved business conditions and hence directly stimulate the U.S. economy. On the other hand, unexpected positive innovations to aggregate demand also increase the real price of crude oil, and indirectly slow down the U.S. economic activity. Results in the middle panel of Figure 1 show that at least in the short-run the stimulating effect prevails on the growth-retarding effect, the net outcome being a decrease in stock price volatility. This result is complementary to the findings of Kilian and Park (2009), who show that an unexpected increase in the global demand for all industrial commodities causes a sustained increase in U.S. stock returns.

Shocks to the physical supply of crude oil, or to oil-specific demand, might indicate a higher degree of macroeconomic uncertainty. We have shown that unexpected increases in the precautionary demand for crude oil significantly increase realized volatility, while realized volatility is unaffected by oil supply shocks. The lack of reaction of stock price volatility to oil supply shocks can be explained in terms of the temporary and limited response of the real price of oil to shocks from the supply side of the market (Kilian, 2009). Therefore, to the extent that shocks to the supply of crude oil do not reduce the long-run profitability of corporate investments, investors' plans will be unaffected (Güntner, 2014).

These findings are in line with those of Kang and Ratti (2013a,b), who report very similar results for an index of policy uncertainty. Compared with Degiannakis et al. (2014), who study the impact of structural oil market shocks on the volatility of European stocks, our analysis leads to different conclusions. These authors show that the impact of oil price shocks due to unanticipated supply reductions or oil-specific demand increases is negligible. While these results can be partially explained by the differences in the fundamentals driving the price of stocks in the U.S. and European markets, the empirical methodology followed by the authors should be also considered.

Specifically, the reduced-form of the VAR of Degiannakis et al. (2014) includes four lags on the same variables used in our study as well as in Kilian (2009), namely oil production and global real economic activity, while the global price of oil is represented by (the nominal log-returns on) the price of Brent. There are at least three points that deserve attention. First, the choice of using Brent instead of the refiner's acquisition cost of imported crude oil as a proxy for the global price of crude oil is questionable, because the price of Brent was not available prior to the 1980s. Moreover, as illustrated by Bastianin et al. (2014) among others, it is not clear a priori whether the price of Brent can serve as a benchmark for the price oil. Second, log-differencing the real price of oil is problematic given the low power of unit root tests and may render their estimates inconsistent. Gospodinov et al. (2014) show that in case of doubt the level specification is more reliable than relying on pre-tests. A third potential pitfall is the use of four lags only. As pointed out by Kilian (2009) and Kang and Ratti (2013a), long lags are important in structural models of the global oil market to account for the low frequency co-movement between the real price of oil and global economic activity. Moreover, when working with monthly data, including less than 12 lags might be problematic if the series are characterized by seasonality (see Günter, 2014). A case in point is the monthly world production time series that the authors use in their model.

**3.2 Does the impact of oil price shocks vary across industries?**

Oil price shocks might have direct input-cost effects: higher energy prices reduce the usage of oil and hence lower the productivity of capital and labor. Alternatively, if higher energy prices lower the disposable income of consumers, the transmission is due to an income effect that reduces the demand for goods. In any case, these alterative channels of transmission suggest that the response of volatility might be different across industries. Heterogeneous responses might depend either on the level of energy intensity, or on the nature of the good produced or service provided.

We focus on the volatility of four industry portfolios selected among the 49 provided by Kenneth French, namely: oil and gas, precious metals, automobile and retail. The shares of firms in the oil and gas and automotive industry should be very sensitive to the price of crude oil. Oil and gas companies have the most energy intensive production processes. The volatility of the shares of auto producers is interesting because car sales and, more generally, the purchase of durable goods might be delayed if oil price is high or expected to be high. The rationale for including the retail industry is that, if an increase in the price of crude oil is passed on to gasoline prices, consumers are bound to devote a larger share of their income to fill up their cars' tanks, therefore they should reduce their spending in other goods. Firms in the precious metal industry have been considered because it is believed that investors will tend to buy more gold and silver (safe-haven assets) when the level of political uncertainty is high. Moreover, the choice of these four industries allows to compare our results with those of Kilian and Park (2009) and Kang and Ratti (2013a).

**[FIGURE 2 ABOUT HERE]**

The first noticeable result from Figure 2 is the shape of the estimated IRFs to any of the three shocks, which is quite similar across industries. On the contrary, the responses change depending on the cause underlying the oil price shock.

Shocks to the supply of crude oil boost the realized volatility of petroleum and natural gas companies on impact, but after one month the response is no longer statistically distinguishable from zero. Six months after an unexpected supply reduction, the response of realized volatility becomes positive and statistically significant at the 68% confidence level in the oil and gas, precious metals and retail industries, but within four months at most it reverts to zero in all cases.

Unexpected increases in the aggregate demand for all industrial commodities yield very similar volatility responses across industries. The realized volatility of all portfolios drops on impact, suggesting that, in the short-run, the direct stimulating effect of aggregated demand shocks dominates the indirect growth-retarding effect of increased crude oil prices. The response is negative and statistically significant based on one-standard error bands for at most a year. The indirect negative impact on the U.S. economy associated to higher oil prices generated by a shock to aggregate demand reaches its maximum after the first year, but the volatility response is positive and statistically significant at 68% and 95% confidence levels only for companies in the automotive industry.

Irrespective of the industry, an unexpected increase in oil-specific demand yields volatility responses that are mostly negative, but not statistically distinguishable from zero on impact. After at least a quarter from the shock, the responses in all industries switch sign from negative to positive. These volatility increases show different levels of persistence and statistical significance depending on the industry, but after a year all of them are not statistically distinguishable from zero.

All in all, these results highlight that the link between volatility responses and energy intensity of the industry is weak at best. As an example, the magnitude and the shape of the responses of the oil and gas portfolio are not very different from those of other, less energy intense, industries.

The finding that the response of shock volatility is homogeneous across different industries is complementary with the results of existing studies, such as Kilian and Park (2009) and Kang and Ratti (2013a), who have analyzed the response of cumulative returns on the same set of portfolios. Their results show that a given shock can have very different impacts on the value of stocks, depending on the industry and on the underlying causes of the oil price increase. One noticeable difference is that our analysis shows that only the origin of the oil price shock matters, whereas the volatility response to the same shock is very similar across industries, although with a different timing. These results suggest that investors and risk managers should be aware of the causes underlying the oil price shock to optimally adjust their portfolios.

# 4. Robustness checks

## 4.1 Alternative oil supply and oil-specific demand shock proxies

We have shown that the volatility of the U.S. stock market has been resilient to oil price increases driven by supply interruptions. Since supply-driven oil price shocks are often seen as the main channel through which the adverse effects of higher energy prices are transmitted to the economy, this result should be subject to additional investigation.

The first robustness check relies on the work of Kilian (2008a), who uses production data for measuring exogenous shocks to the supply of crude oil due to geo-political events in the OPEC countries.

We have updated Kilian's (2008a) measure of exogenous oil supply shocks to include the crude oil production shortfall as a consequence of the Libyan Civil War started in February 2011. The original series ends in September 2004, while the updated measure spans January 1973 until December 2013.

We have calculated a counterfactual production level for each OPEC country where a political event - such as a war - has caused a shortfall in crude oil production. The production shortfall is defined as the difference between a country's actual crude oil production and the counterfactual production level. The latter is the level of crude oil production that would have prevailed in the absence of the exogenous events which are responsible for the shortfall. It is obtained by extrapolating the pre-event production level based on the average growth rate of production from those countries which are not affected by the events.[8] Exogenous crude oil production shortfalls are then aggregated, expressed in percentage of world crude oil production and first-differenced.

---

[8] The construction of counterfactual production levels follows the details in Kilian (2008a). The dates of exogenous events and the list of countries used to calculate the counterfactual are the shown in Kilian's (2008c) Tables 1 and 2. We have updated the content of these tables so as to accommodate the exogenous production shock due to the Libyan Civil War and the new

The updated Kilian's (2008a) measure of exogenous oil production shocks is shown in Figure 3.[9] The counterfactual for Libya starts in February 2011 and is based on the average growth rate of production in Algeria, Angola, Ecuador, Nigeria, Qatar, United Arab Emirates. As of March 2011, Kilian's measure of production shocks indicates that the Libyan Civil War has led to a production shortfall representing approximately one percent of world crude oil production. Comparison with production shortfalls caused by other exogenous events shows that the Libyan events have caused the smallest production disruption among those considered by Kilian (2008a).

As shown in the leftmost panel of Figure 4, the response of volatility to this alternative oil supply shock proxy is not statistically distinguishable from zero.[10] This is consistent with the results

---

composition of OPEC. As of June 2015 OPEC members are: Algeria, Angola, Ecuador, Iran, Iraq, Kuwait, Libya, Nigeria, Qatar, Saudi Arabia, United Arab Emirates and Venezuela. Production data have been sourced from the Energy Information Administration's website.

[9] Although the measure of exogenous oil production shocks shown in Figure 3 is visually identical to the original shown in Kilian's (2008a) Figure 7, there might be small numerical differences between the original and updated series due to the fact that the list of OPEC member countries changed after the first draft of the Kilian paper was written. Actually, neither Angola, that joined OPEC in 2007, nor Ecuador, that suspended its membership from December 1992 until October 2007, were considered in Kilian (2008a). As pointed out by Alquist and Coibion (2014), notwithstanding these numerical discrepancies, the correlation between the original Kilian's (2008a) measure and our updated series is 0.99.

[10] Results in both panels of Figure 4 are based on bivariate structural VAR models of order 12, with the shock ordered first and *RV* ordered last. Specifically, in the leftmost panel the exogenous OPEC oil supply shock proposed by Kilian (2008a) is ordered first and *RV* is ordered last, while in the rightmost panel the shock based on the share of respondents to the University of Michigan Survey of Consumer Sentiment who quote gasoline shortages as a relevant motivation to postpone the

in the leftmost panel of Figure 1: shocks to the supply of crude oil cannot be held responsible for volatility increases.

**[FIGURES 3, 4 ABOUT HERE]**

As a second robustness check, we consider an alternative measure for the oil market-specific shock. Following Ramey and Vine (2010), we use the proportion of respondents to the University of Michigan's Survey of Consumer Sentiment, who cite the price of gasoline, or possible fuel shortages, as a reason for poor car-buying conditions. The rightmost graph in Figure 4 shows that the volatility response estimated with this alternative proxy is very similar to what obtained when considering shocks to the precautionary demand for crude oil (see the rightmost panel in Figure 1).

### 4.2 Alternative sampling frequency and volatility proxies

Three additional robustness checks involve the use of the logarithm of *RV* in place of *RV*, a different data sampling frequency, as well as alternative volatility proxies. These results are not shown here, but are available on request.

Since aggregate stock return volatility is positively skewed and leptokurtic, researchers often use the logarithm of realized volatility (see Andersen et al., 2001). Therefore, we have re-estimated the structural VAR model considered in Section 3.1 with the logarithm of *RV* in place of the *RV* of the S&P500 index. Our results show that our main conclusions are not affected when considering the log-transformed *RV*.

All results presented so far rely on structural VAR models estimated on monthly variables. However, as pointed out by Baumeister and Kilian (2014), macroeconomic models used in central banks are often specified at quarterly frequency. For this reason, we now consider a structural VAR

---

purchase of a car is ordered first and *RV* is ordered last. This identification scheme is consistent with the assumption that the real price of crude oil is predetermined with respect to macroeconomic and financial variables.

model of order 8 with variables sampled at quarterly frequency. Results are qualitatively identical to those in Figure 1 for the monthly structural VAR of order 24.

The last robustness check we consider uses the conditional volatility from a Generalized Autoregressive Conditional Heteroskedasticity model, GARCH(1,1), and the Chicago Board Options Exchange (CBOE) volatility index (VIX) as alternative measures of the S&P500 stock index volatility. Our results do not change when these alternative proxies replace *RV*. In other words, the response of volatility to oil price shocks depends on their causes, while it is not affected by the choice of the volatility proxy.

**5. Conclusions**

Stock market volatility and the price of crude oil, being two of the variables that policy makers and financial investors track most closely (see e.g. Bernanke, 2006; Brown and Sarkozy, 2009), are often front page news. Moreover, academic research has analyzed in detail the effects of oil price shocks on macroeconomic and financial variables.

In this paper we have shown that, in order to understand the response of the U.S. stock market volatility to changes in the price of crude oil, the causes underlying oil price shocks should be disentangled. This conclusion has been extended to the analysis of the impacts of oil price shocks on different industry portfolios. Contrary to what expected, the impact of supply shortfalls is negligible and volatility responds mostly to shocks hitting aggregate and oil-specific demand. Evidence of heterogeneous volatility responses across industries is modest at best.

The result that stock volatility reacts differently to shocks originating from the supply and demand side of the crude oil market has important implications for policy makers, investors, risk managers, asset allocation strategists and macroeconomic model builders. For instance, studies on the relation between monetary policy and asset price volatility (e.g. Bernanke and Gentler, 1999), should be extended to include the global oil market, building on the global DSGE framework developed in Bodenstein, Guerrieri and Kilian (2012), for example.

**Figure 1. Responses of S&P500 volatility to structural oil market shocks**

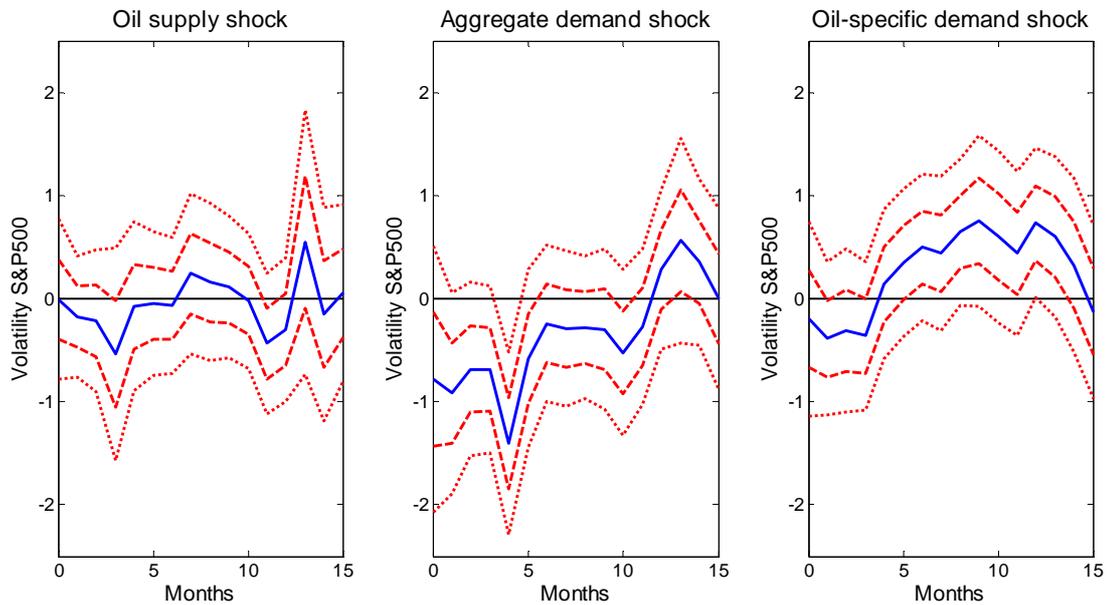

*Notes*: each panel shows the response of the annualized realized standard deviation of the S&P500 index to a one-standard deviation structural shock (continuous line), as well as one- (dashed line) and two-standard error bands (dotted line). Estimates are based on a recursively identified four-variable structural VAR model of order 24. Confidence bands (at 68% and 95% levels) are based on a recursive-design wild bootstrap with 2000 replications (see Gonçalves and Kilian 2004).

**Figure 2. Responses of industry portfolios volatility to structural oil market shocks**

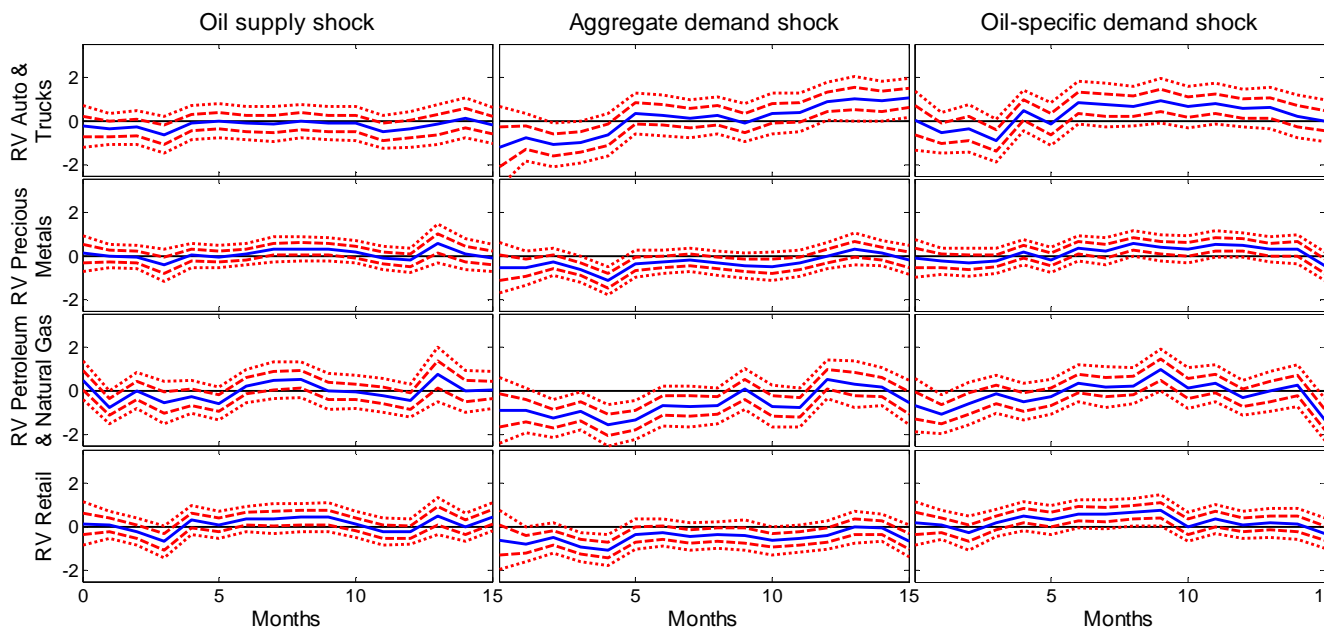

*Notes*: each row of the figure shows the response of the annualized realized standard deviation of the industry portfolio indicated on the label of the vertical axis to a one-standard deviation structural shock (continuous line), as well as one- (dashed line) and two-standard error bands (dotted line). For each industry portfolio estimates are based on a recursively identified four-variable structural VAR model of order 24. Confidence bands (at 68% and 95% levels) are based on a recursive-design wild bootstrap with 2000 replications (see Gonçalves and Kilian 2004).

**Figure 3. Exogenous oil supply shocks (January 1973 – December 2013)**

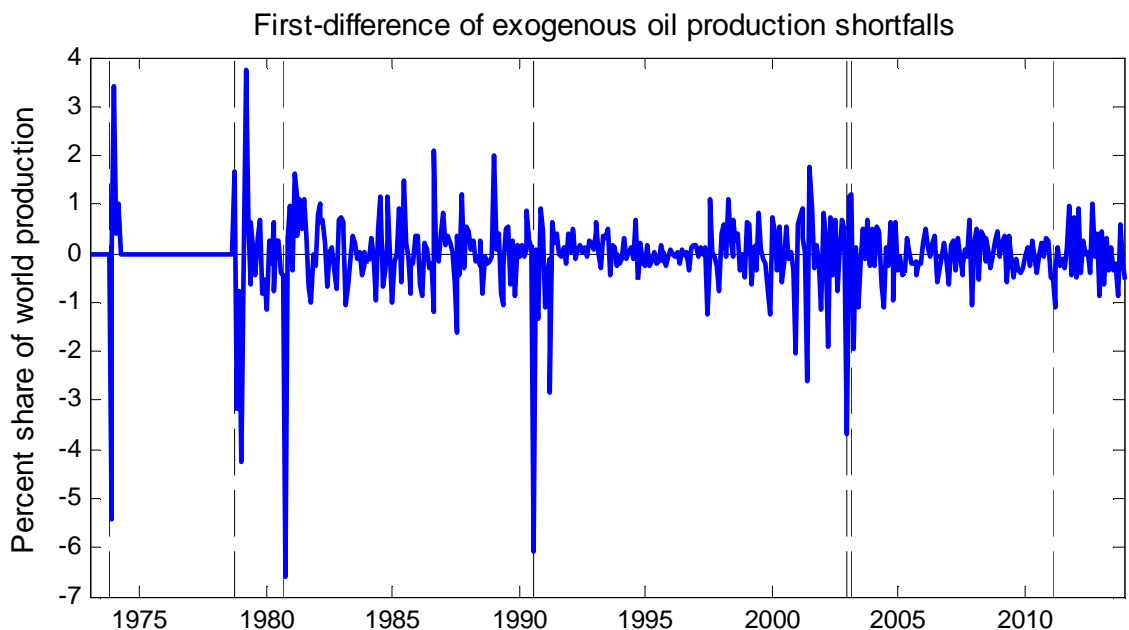

*Notes:* the figure shows the measure of exogenous oil supply shocks due to Kilian (2008a). The first five vertical dashed lines identify key oil dates shown in Table 1 of Kilian (2008c), while the last line is drawn in correspondence of February 2011, beginning of the Libyan Civil War. The remaining key oil dates are: October 1973 (Yom Kippur War and Arab oil embargo), October 1978 (Iranian revolution), September 1980 (Iran-Iraq War), August 1990 (Persian Gulf War), December 2002 (Civil unrests in Venezuela) and March 2003 (Iraq War).

**Figure 4. Responses of S&P500 volatility to exogenous oil supply shocks and gas-shortages**

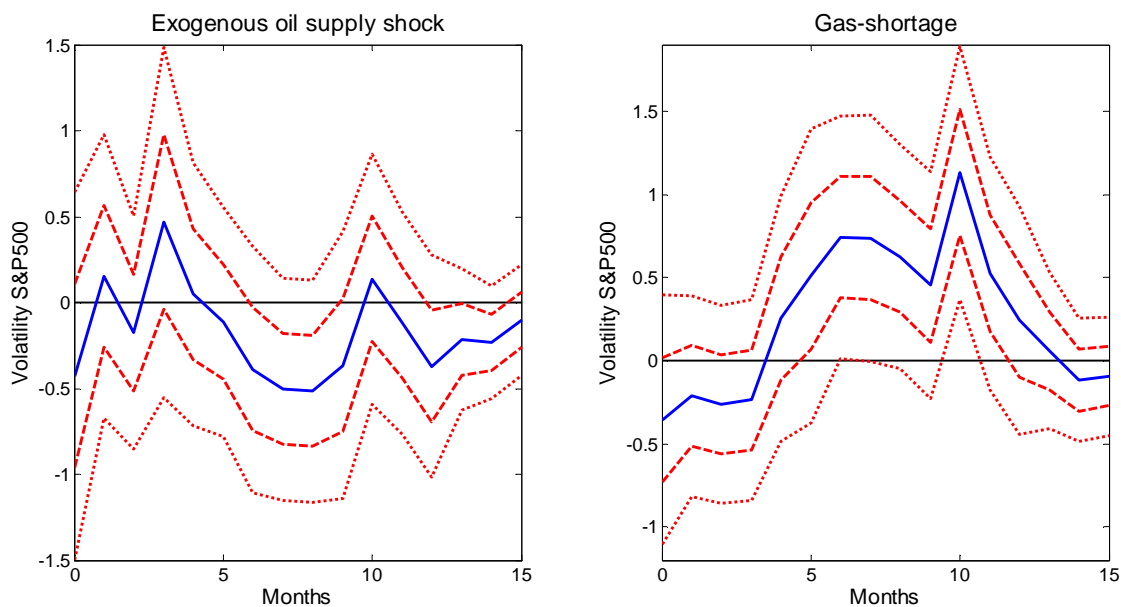

*Notes*: each panel shows the response of the annualized realized standard deviation of the S&P500 index to a one-standard deviation structural shock (continuous line), as well as one (dashed line) and two-standard error bands (dotted line). Confidence bands (at 68% and 95% levels) are based on a recursive-design wild bootstrap with 2000 replications (see Gonçalves and Kilian 2004). Estimates are based on bivariate structural VAR models of order 12, with the shock ordered first and the volatility series ordered last. In the leftmost panel the shock is the measure of the exogenous OPEC oil supply shock proposed by Kilian (2008a), while in the rightmost panel the shock is based on the (percent change of the) share of respondents to the University of Michigan Survey of Consumer Sentiment, who quote gasoline shortages as a relevant motivation to postpone the purchase of a car.